# 3-μm Reflectance Spectroscopy of Carbonaceous Chondrites under Asteroid-like Conditions

Driss Takir[1,*]; Karen R. Stockstill-Cahill[2]; Charles A. Hibbitts[2], Yusuke Nakauchi[3]

[1]Jacobs/ARES, NASA Johnson Space Center, Houston TX 77058-3696, USA

[2]Johns Hopkins University Applied Physics Laboratory, Laurel, Maryland 20273, USA

[3]JAXA Institute of Space and Astronautical Science, Sagamihara, Japan

***Corresponding author email:** driss.takir@nasa.gov

**Editorial correspondence to:**

Driss Takir

Astromaterials Research and Exploration Science Division

NASA Johnson Space Center

2101 NASA Parkway

Building 36, Room 126

Houston TX 77058-3696

Pages: 24
Tables: 2
Figures: 27

**Abstract**

We measured 3-μm reflectance spectra of 21 meteorites that represent all carbonaceous chondrite types available in terrestrial meteorite collections. The measurements were conducted at the Laboratory for Spectroscopy under Planetary Environmental Conditions (LabSPEC) at the Johns Hopkins University Applied Physics Laboratory (JHU APL) under vacuum and thermally-desiccated conditions (asteroid-like conditions). This is the most comprehensive 3-μm dataset of carbonaceous chondrites ever acquired in environments similar to the ones experienced by asteroids. The 3-μm reflectance spectra are extremely important for direct comparisons with and appropriate interpretations of reflectance data from ground-based telescopic and spacecraft observations of asteroids. We found good agreement between 3-μm spectral characteristics of carbonaceous chondrites and carbonaceous chondrite classifications. The 3-μm band is diverse, indicative of varying composition, thus suggesting that these carbonaceous chondrites experienced distinct parent body aqueous alteration and metamorphism environments. The spectra of CI chondrites, from which significant amount of water adsorbed under ambient conditions was removed, are consistent with Mg-serpentine and clay minerals. The high abundances of organics in CI chondrites is also associated with the mineralogy of these chondrites, oxyhydroxides- and complex clay minerals-rich. CM chondrites, which are cronstedtite-rich, have shallower 3-μm band than CI chondrites, suggesting they experienced less aqueous alteration. CR chondrites showed moderate aqueous alteration relative to CI and CM chondrites. CV chondrites, except for Efremovka, have a very shallow 3-μm band, consistent with their lower phyllosilicate proportions. CO chondrites, like most CVs, have a very shallow 3-μm band that suggest they experienced minor aqueous alteration. The 3-μm band in CH/CBb is deep and broad centered ~ 3.11 μm, possibly due to the high abundance of FeNi metal and presence of heavily hydrated clasts in these chondrites. The 3-μm spectra of Essebi (C2-ung) and EET 83226 are more consistent with CM chondrites' spectra. The 3-μm spectra of Tagish lake (C2-ung), on the other hand, are consistent with CI chondrites. None of these spectral details could have been resolved without removing the adsorbed water before acquiring spectra.

**Introduction**

This investigation completes work begun under Takir et al. (2013) that explored 10 samples of CM and CI carbonaceous chondrite types. In this paper we expand that study, investigating 21 more samples that represent all carbonaceous chondrite types available in terrestrial meteorite collections, including CI, CM, CR, CV, CO, CH, CB, CK, and C2-ung chondrites. With this complete investigation of the 3-μm band under vacuum and thermally-desiccated conditions (asteroid-like conditions), we can better constrain the mineralogy and understand the association of hydrated minerals with organics (and other chemical compounds) on the surface of carbonaceous asteroids, and therefore provide valuable constraints on the current dynamical and thermal theories of the formation and evolution of the early solar system. Most previous spectroscopic studies of meteorites focused on the 0.35-2.5 μm spectral region, which is not strongly affected by atmospheric water contamination (e.g., Johnson and Fanale 1973, Gaffey 1974, Salisbury et al. 1975, Cloutis et al.2011a, 2012b, 2012a, 2012b, 2012c, 2012d, 2012e, 2012f). Here we present the first comprehensive 3-μm dataset of carbonaceous chondrites that includes important absorption features (e.g., hydroxyl, organics, carbonates) and requires that the spectroscopic measurements to be done in asteroid-like conditions for direct comparison with ground- and space-based observations.

Carbonaceous chondrites provide important clues about the formation and evolution of the early solar system because they are widely thought to be unaltered by extensive melting and differentiation of their parent bodies (Van Schmus and Wood 1967). Remote sensing is considered a powerful tool for linking carbonaceous chondrites with their possible asteroid parent bodies, providing more information on the petrological and geochemical environments in which these asteroids were formed. JAXA's Hayabusa2 and NASA's OSIRIS-REx spacecraft have rendezvoused with asteroids (162173) Ryugu (Watanable et al. 2019) and (10955) Bennu (Lauretta et al. 2018), respectively, both are thought to be carbonaceous asteroids and are expected to contain water and/or organics. These two missions confirmed the presence of hydrated minerals on these two carbonaceous asteroids.

Since it is not possible to return samples to Earth from every small body in the solar system, the information obtained from remote sensing data (ground- or space-based) is important to characterize the surface composition of these bodies. Studying carbonaceous chondrites is also important for sampling and landing site characterization and selections for the few sample return missions and human exploration missions. Furthermore, the study of carbonaceous chondrites will help place the analysis of the returned samples from carbonaceous small bodies in a broader 'solar system' context.

The integrated area of the 3-µm band is sensitive to the amount of hydrated minerals and/or water content in carbonaceous chondrites (e.g., Miyamoto and Zolensky 1994). Previous studies used reflectance spectra of carbonaceous chondrites measured in ambient conditions to derive 3-µm spectral parameters, such as, band center, band depth, and band area, which were later used to estimate the H/Si ratio for asteroids (e.g., Sato et al. 1997). Takir et al. (2013) using meteorites and Milliken and Mustard (2007) using clays spectra showed, however, that ambient conditions can affect the calculation of spectral parameters of the 3-µm band, and hence the interpretation of remote sensing spectra around this band, due to the contamination by significant amounts of adsorbed atmospheric water. In this study we measured infrared (IR) reflectance spectra of carbonaceous chondrites in vacuum and room temperature after overnight heating at T ~ 375 K to minimize the presence of adsorbed atmospheric water. The relatively low temperature at which we heated the samples, avoided mineral alteration (e.g., Milliken and Mustard 2007) that could otherwise be relevant to carbonaceous chondrites at T > 400 K (e.g., Tonui et al. 2011).

**Methodology**

*1. Samples*

We carefully selected our meteorite samples for this study, all of them are falls except for five meteorites (Table 1). MIL15328, DOM 10085, and EET83226 were recovered from Antarctica, with weathering grades of B, C, and A/B, respectively. NWA 4964 and SAU 290 are desert meteorites that were recovered in Morocco and Oman, respectively, with a moderate weathering grade. We ground meteorite chips (100-150 mg) into fine powders (~100 µm), using a dry ceramic mortar and pestle. Because we got the samples as a loan and were limited by the sample sizes, we could not sieve the meteorite samples and precisely measure their grain size distributions.

**Table 1**. We measured 3-µm spectra of 21 carbonaceous chondrites in vacuum and thermally-desiccated, covering all carbonaceous chondrite types available in terrestrial meteorite collections.

| Meteorite* | Type/Weathering grade | Meteorite* | Type/Weathering grade |
|---|---|---|---|
| Alais | CI1 | Mokoia | CV3 |
| Orgueil | CI1 | Vigarano | CV3 |
| Murchison | CM2 | Felix | CO3.3 |
| Crescent | CM2 | Warrenton | CO3.7 |
| Banten | CM2 | NWA 4964** | CK3/B |
| Murray | CM 2 | SAU 290** | CH3/B |
| Al Rais | CR2 | Isheyevo | CH/CBb |
| MIL 15328** | CR2/B | EET 83226** | C2-ung/A/B |
| DOM 10085** | CR2/C | Essebi | C2-ung |
| Allende | CV3 | Tagish Lake | C2-ung |
| Efremovka | CV3 | | |

*Meteorite information from the Meteoritical Bulletin Database. (https://www.lpi.usra.edu/meteor/). A = minor rustiness; B = moderate rustiness; C = major rustiness. **These meteorites are finds.

*2. Spectroscopic laboratory measurements procedure*

We measured bidirectional reflectance spectra (*incidence* = 15°, *emission* = 45°, *phase angle* = 60°) of carbonaceous chondrites at the Laboratory for Spectroscopy under Planetary Environmental Conditions (LabSPEC) at the Johns Hopkins University Applied Physics Laboratory (JHU APL) under vacuum- and thermally-desiccated conditions, following the methodology used in Takir et al. (2013). Spectra were measured from ~0.8 to ~ 8 µm using a high-vacuum chamber system ($10^{-6}$ to $10^{-7}$ torr) with a Bruker Vertex 70 FTIR spectrometer and an external liquid-nitrogen cooled MCT detector. Samples were placed in the copper sample holder and retained in place with a 1 mm-thick $MgF_2$ window. A diffuse gold plate mounted immediately below the sample holder was used as the IR reflectance standard, and a thermocouple embedded in the sample provided an accurate temperature measurement. After placing the sample in the holder and securing the window over the sample, the holder was installed onto the cryostat inside the chamber. After installing the holder with sample in it into the chamber, a spectrum is obtained under 'ambient' conditions, with the sample exposed to the laboratory environment. Next, the chamber was sealed and evacuated, and spectra of the initially desiccated sample were obtained. However, much adsorbed water generally remained in samples, so they were baked out. While under vacuum, a resistive heater warmed the back of the copper holder and thus the sample for removing adsorbed water. Sample temperature reached ~ 375K (which is consistent with the warmest temperature near-Earth and Main Belt asteroids experience) for ~ 10 to 12 hours overnight (Delbo and Michel 2011). Then in the following day, the heater was turned off, and spectra were taken again under vacuum after the sample had cooled.

*3. Spectral analysis*

For our spectral analysis and band parameters calculations, we used a standard technique described in Cloutis et al. (1986). The 3-µm band was isolated and then divided by a straight-line continuum, which was determined by two reflectance maxima just shortward of the absorption 2.60-2.65 µm and just longward of the absorption 3.5-3.85 µm. To be consistent with the asteroid spectral analysis (Takir and Emery 2012, Takir et al. 2015), the band depths for chondrites spectra were computed at ~2.90 µm relative to the linear continuum that was calculated from a linear regression of data from 1.95-2.5 µm. This formulation (K-band) is defined as:

$$D_{2.90} = 1 - (R_{2.90} / R_c) \quad (1),$$

where $R_{2.90}$ is the reflectance at ~2.90 µm, and $R_c$ is the reflectance of the continuum at the same wavelength as $R_{2.90}$ as interpolated via the calculated linear continuum (Figure 1).

We then calculated the band areas by integrating the spectral curve below the straight-line:

$$\text{Band Area} = \int_B^A \frac{R(\lambda)}{R_c(\lambda)} \, d\lambda \quad (2).$$

We used an average of five measurements of band maxima determined by choosing band continua A: 2.60-2.65 µm and B: 3.5-3.85 µm. The band centers were calculated by applying a sixth-order polynomial fit to the central part of 3-µm band minima and finding the wavelength corresponding to the minimum reflectance for that fit. Uncertainty for band area and center measurements is expressed as a 2σ standard deviation of the fit of the function to the data. For spectra with sufficiently well-defined organics absorption around 3.4-3.5 µm, we subtracted those organics band area from the overall 3-µm band area to improve the accuracy of the measurements of the integrated band area.

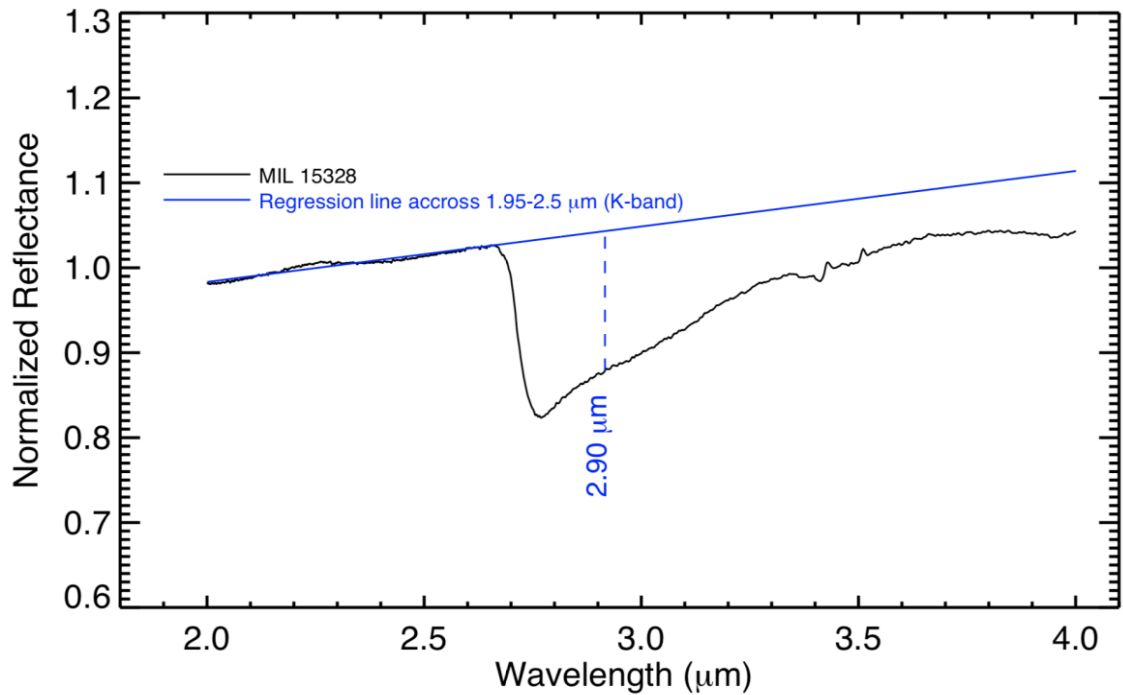

**Figure 1.** The 2.9-µm band depth was calculated relative to the continuum, the linear regression line across the 1.95-2.5 µm spectral region (K-band). This method is consistent with the methods used for ground-based spectra of asteroids (Takir and Emery 2012, Takir et al. 2015).

**Results**

Figure 2 shows IR reflectance spectra of 21 carbonaceous chondrites included in this study. Two CI chondrites (Alais, Orgueil) are shown in Figures 2a-b, spectra of four CM chondrites (Murchison, Crescent, Banten, and Murray) in Figures 2c-f, spectra of three CR chondrites (Al Rais, MIL 15328, DOM 10085) in Figures 2g-i, spectra of four CV chondrites (Allende, Efremovka, Mokoia, Vigarano) in Figures 2j-m, spectra of two CO chondrites (Felix, Warrenton) in Figures 2n-o, spectrum of one CK chondrite (NWA 4964) in Figure 2p, spectra of two CH/CB chondrites (SAU 290, Isheyevo) in Figures 2q-r , and spectra of three C2-ung chondrites (Tagish lake, EET 83226, Essebi) in Figures 2s-u. All spectra were normalized to unity at 2.2 µm and represent measurements that were conducted at ambient conditions (gray curves) and measurements conducted in vacuum and thermally-desiccated (black curves). Table 2 shows the 3-µm band parameters (band center, depth, and area), derived using the techniques discussed in the methodology section, for the 21 carbonaceous chondrites spectra measured in vacuum and thermally-desiccated conditions in this study.

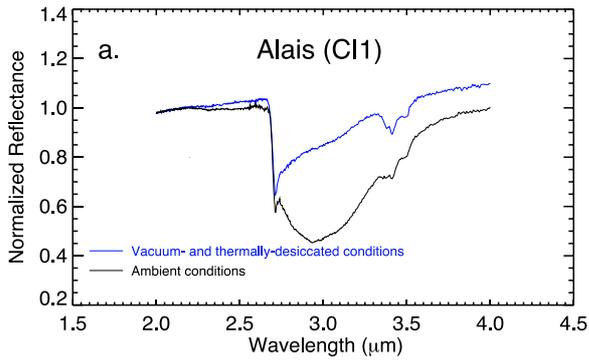
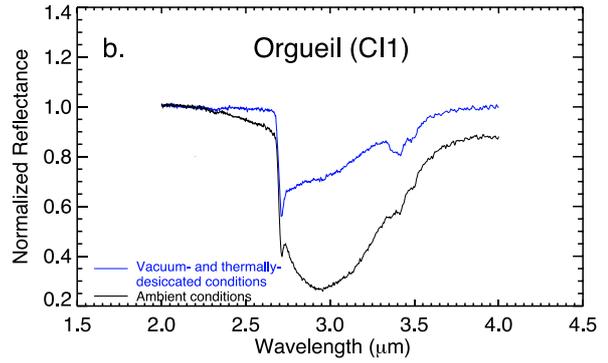
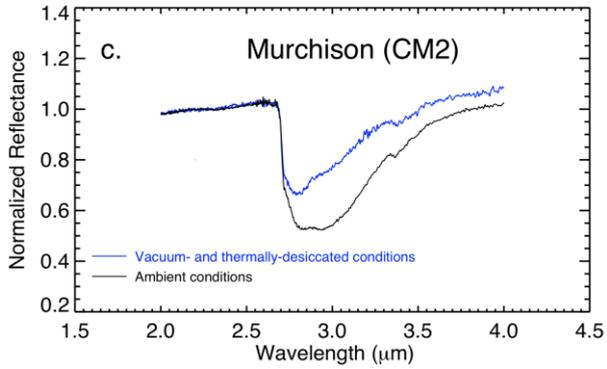
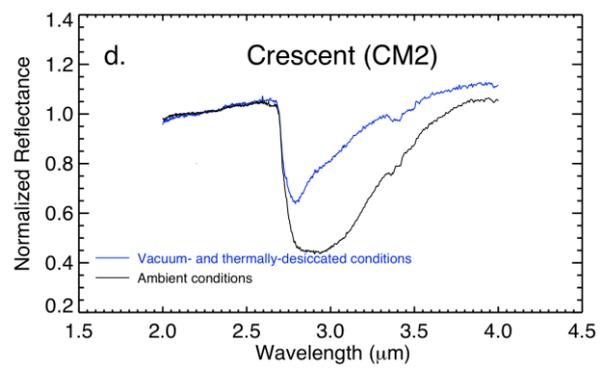
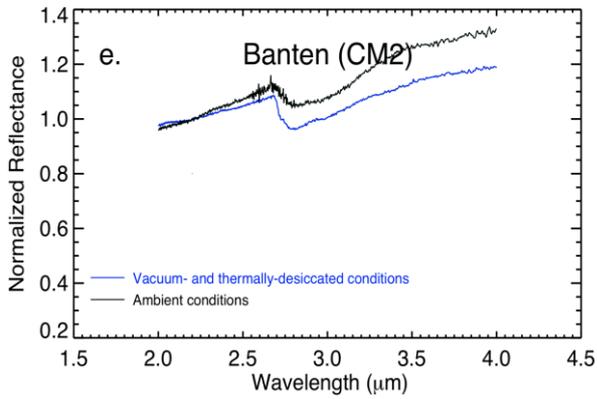
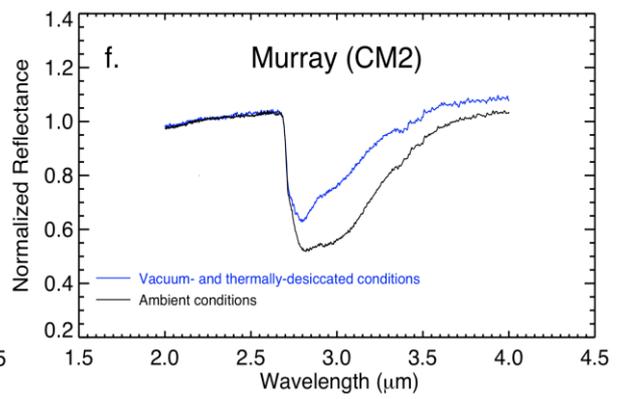
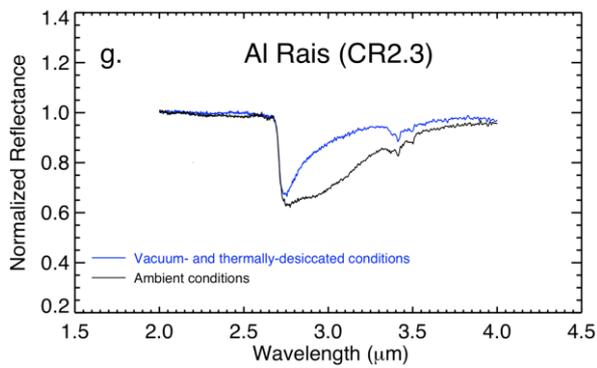
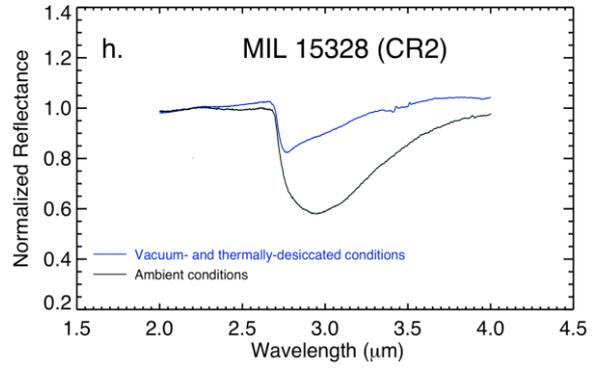

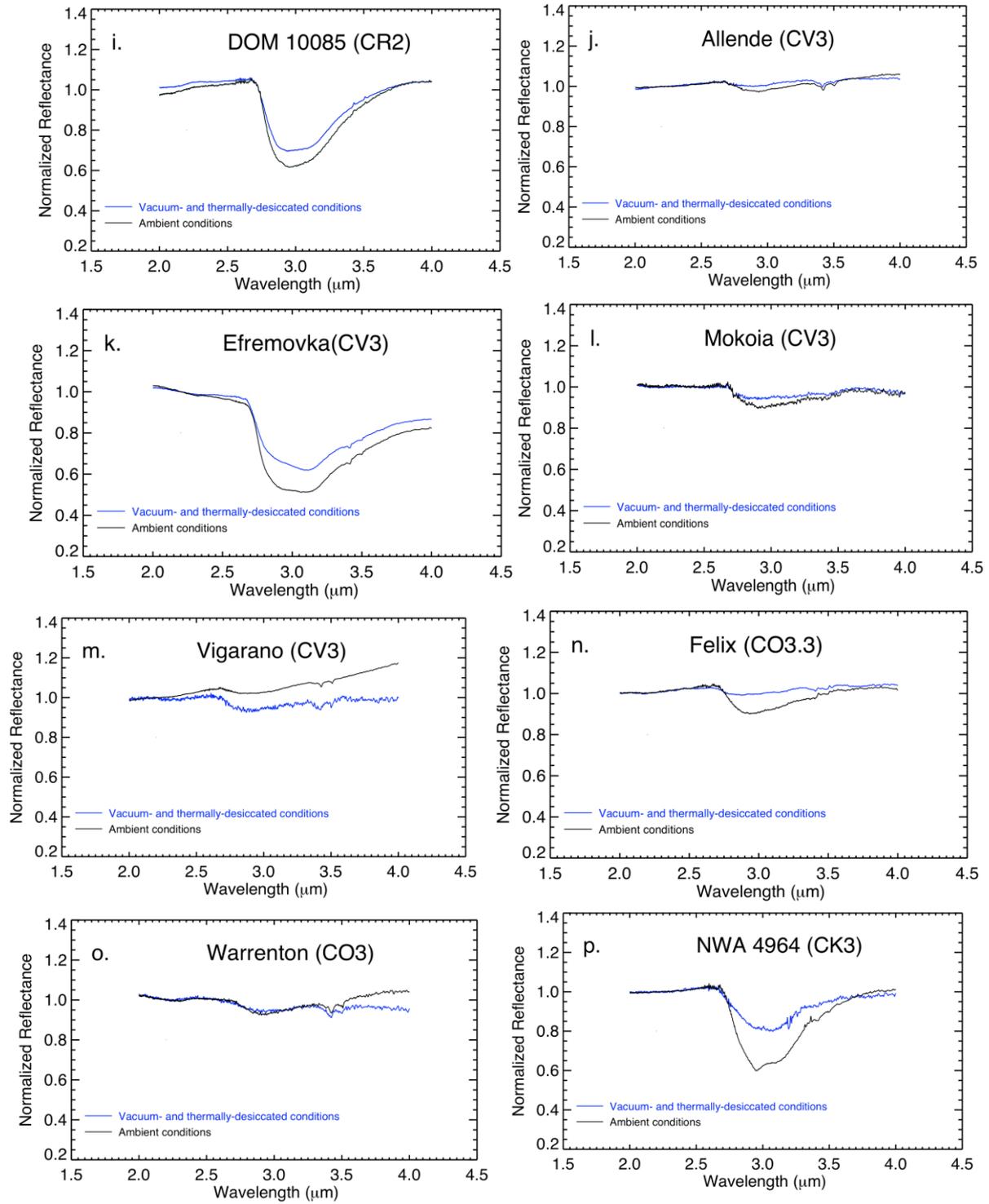

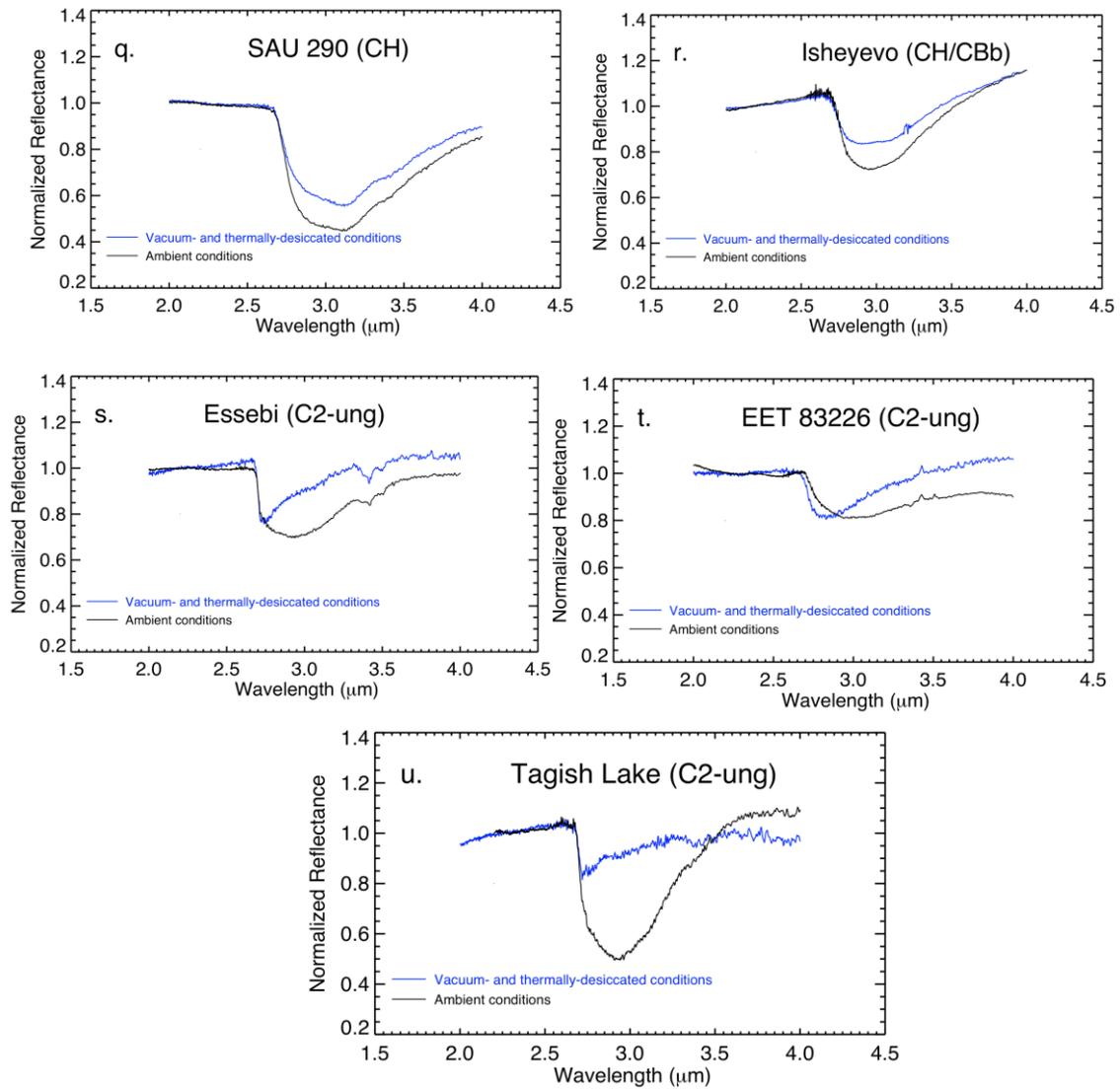

**Figure 2**. a-u. Average IR reflectance spectra of CI, CM, CR, CV, CO, CK, CH, CB, and C2-ung chondrites measured at ambient (blue curves), and asteroid-like conditions (vacuum and thermally-desiccated conditions) (black curves). All spectra have been normalized to unity at 2.2 μm.

**Table 2.** 3-μm band parameters for carbonaceous chondrite spectra measured in asteroid-like (vacuum and thermally-desiccated) conditions.

| Meteorite* | Figure 2 | 3- μm band center (μm) | 2.9- μm band depth (%) | 3- μm band area (μm$^{-1}$) |
|---|---|---|---|---|
| Alais | a | 2.714±0.007 | 21.73 | 0.119±0.004 |
| Orgueil | b | 2.713±0.001 | 27.80 | 0.169±0.004 |
| Murchison | c | 2.796±0.004 | 29.35 | 0.152±0.016 |
| Crescent | d | 2.792±0.003 | 26.65 | 0.131±0.002 |
| Banten | e | 2.799±0.003 | 9.65 | 0.051±0.002 |
| Murray | f | 2.797±0.003 | 30.66 | 0.162±0.006 |
| Al Rais | g | 2.745±0.004 | 15.07 | 0.090±0.002 |
| MIL 15328 | h | 2.768±0.001 | 33.51 | 0.196±0.005 |
| DOM 10085 | i | 2.949±0.005 | 15.76 | 0.075±0.002 |
| Allende | j | 2.898±0.016 | 3.62 | 0.010±0.001 |
| Efremovka | k | 3.111±0.007 | 29.63 | 0.165±0.004 |
| Mokoia | l | 2.893±0.022 | 5.20 | 0.032±0.002 |
| Vigarano | m | 2.887±0.019 | 6.14 | 0.039±0.003 |
| Felix | n | 2.881±0.006 | 4.12 | 0.017±0.004 |
| Warrenton | o | 2.923±0.039 | 4.56 | 0.024±0.001 |
| NWA 4964 | p | 3.064±0.015 | 19.53 | 0.098±0.007 |
| SAU 290 | q | 3.108±0.016 | 37.54 | 0.302±0.006 |
| Isheyevo | r | 2.905±0.015 | 20.88 | 0.110±0.009 |
| EET 83226 | s | 2.834±0.015 | 16.39 | 0.085±0.010 |
| Essebi | t | 2.736±0.008 | 15.79 | 0.087±0.006 |
| Tagish Lake | u | 2.723 0.004 | 16.79 | 0.061±0.006 |

**Discussion**

*1. Importance of measuring 3-μm reflectance spectra of carbonaceous chondrites under asteroid-like conditions*

In this work, we focused on the 2-4 μm spectral region that exhibits important diagnostic absorption features, including the water and organics features. Comparisons with reflectance spectra of asteroids that have being measured in space require that we measure meteorite reflectance spectra under asteroid-like conditions in the laboratory. At ambient conditions (room temperature and pressure), the 2-4 μm spectral region is significantly contaminated with adsorbed atmospheric water, which severely affect the calculations of features' spectral parameters (e.g., band center, band shape, band area), and hence, the mineralogical and chemical interpretation of meteorite spectra and their comparisons with asteroid spectra. In some meteorite spectra, adsorbed water can spectrally mask absorption features (e.g., organics, carbonates). Figure 3 shows IR reflectance spectra of a carbonaceous chondrite, Alais, measured at ambient temperature and pressure (gray curve), in vacuum under ambient temperature (dash dot black curve), and in vacuum and slightly elevated temperature at 375K (solid black curve). This figure clearly demonstrates the effects of removing adsorbed water on the spectral parameters. For example, the 3-μm band area decreased from BA=0.341 μm$^{-1}$ in ambient conditions to BA=0.119 μm$^{-1}$ in asteroid-like conditions. In addition, the organic absorption (~3.4-3.5 μm) increased from BA=0.007 μm$^{-1}$ in ambient conditions to BA=0.010 μm$^{-1}$ in asteroid-like conditions.

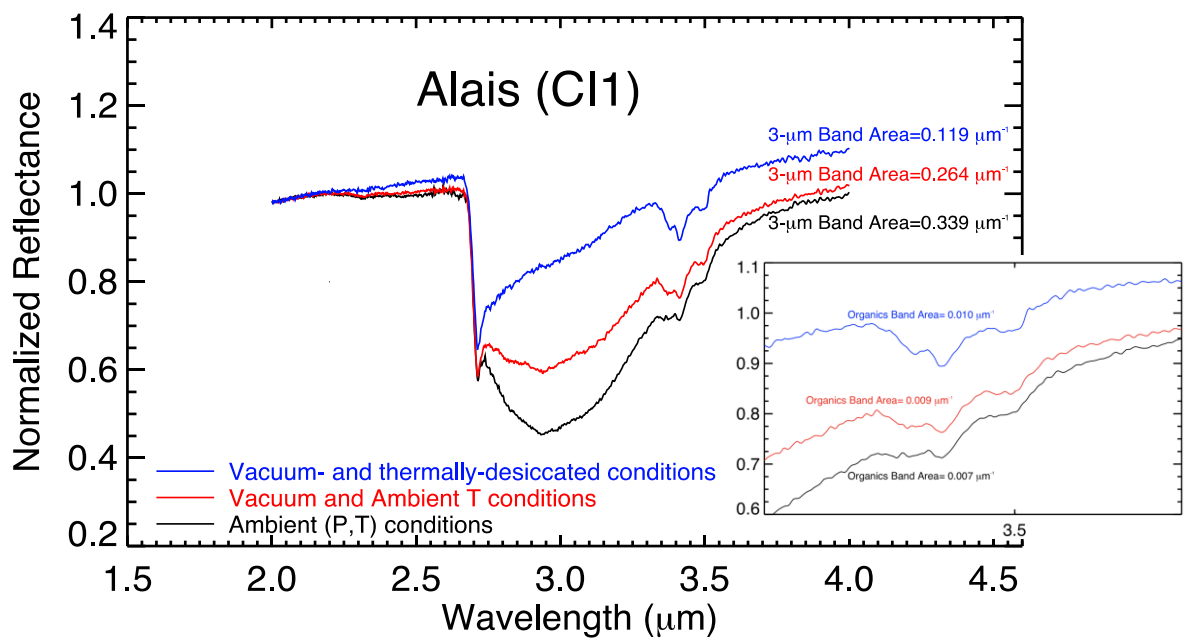

**Figure 3.** IR spectra of a carbonaceous chondrite, CI Alais, measured under ambient pressure and temperature (solid gray curve), immediately after exposure to vacuum and while at ambient temperature (dash dot black curve), and under asteroid-like (vacuum and thermally-desiccated) conditions (solid black curve). The shape and intensity of the 3-μm band, including the organics features around 3.4-3.5 μm, is enormously affected by adsorbed water due to the atmospheric contamination in laboratory at ambient conditions. In vacuum and at T = 375K, the organics features become more pronounced and intense. Note the significant spectral effect of adsorbed

water in the sample measured under ambient conditions.

In general, meteorite spectra show three fundamental vibrations of the water molecule that occur near 3 μm ($v_1$ symmetric and $v_3$ asymmetric OH stretches) and at ~6 μm ($v_2$ H-O-H bend). The 3-um band is sensitive to both hydroxyl and molecular water, whereas the 6-μm band is only affected by molecular water. Both bands are significantly affected by atmospheric (adsorbed) water. To ensure that most or all of adsorbed water is removed from our samples, we heated the sample until the 6-μm band went away, and so $v_3$ asymmetric OH stretch that is due to water contamination in the laboratory (Figure 4). The fact that heating to only 375K was sufficient to remove all molecular water, implies the molecular water is adsorbed and not bound into the mineral structure and also the ease of desiccation suggests all molecular water is likely terrestrial contamination. Any structural water (i.e., hydroxyl groups) in meteorites is not affected at these temperatures (Takir et al. 2013).

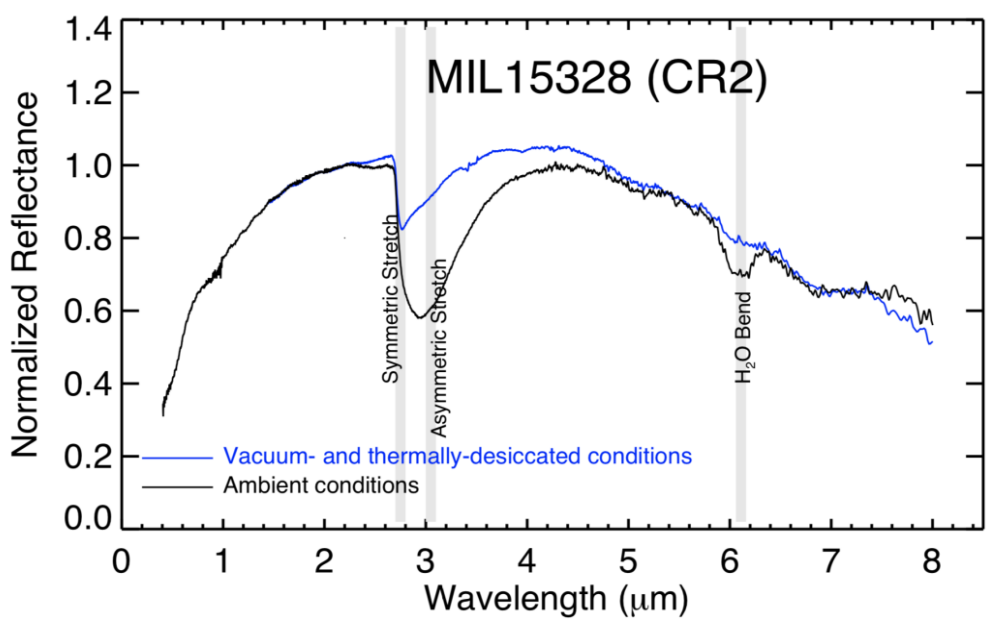

**Figure 4**. IR reflectance spectra of MIL 15328 measured at JHU APL's LabSPEC, covering the 0.8-8 μm spectral range. There is a significant difference (in the 3-μm and 6-μm water bands) between the spectrum that was measured under ambient conditions (gray curve) and asteroid-like (vacuum and thermally-desiccated) conditions (black curve). For direct comparison with ground- and space-based data, meteorite spectra need to be measured in asteroid-like conditions. For confirming there is no residual molecular water ($H_2O$), the 6-μm band needs to be measured.

*2. Diversity in the 3-μm band in carbonaceous chondrites*

In this work we included 21 chondrites that represent all carbonaceous chondrite types available in terrestrial meteorite collections. We found that the 3-μm band is diverse for these samples, suggesting their parent bodies experienced distinct aqueous alteration environments (Figure 5). We derived a very good correlation ($R^2$=0.91) between the band area and the band depth at 2.90 μm (Figure 6). However, we did not find a very good correlation between the 3-μm band shape and carbonaceous chondrite types.

In general, we found a good and qualitative agreement with the petrological classification and the degree of aqueous alteration of carbonaceous chondrites and their 3-μm band parameters (e.g., band area, band center). Overall, as the aqueous alteration increases the 3-μm band becomes deeper and the band center shifts to longer wavelengths. Figure 5 shows this study's spectra in addition to Takir et al. (2013)'s spectra. MIL 00740, a CM2 chondrite with the shallowest and narrowest 3-μm band, is an exception as shown in Figure 5. This chondrite is one of the rare CM chondrites that show pronounced and well-defined olivine and pyroxene absorptions features, indicating it has a high fraction of anhydrous silicates and suggesting it experienced a minor degree of aqueous alteration (Takir et al. 2013).

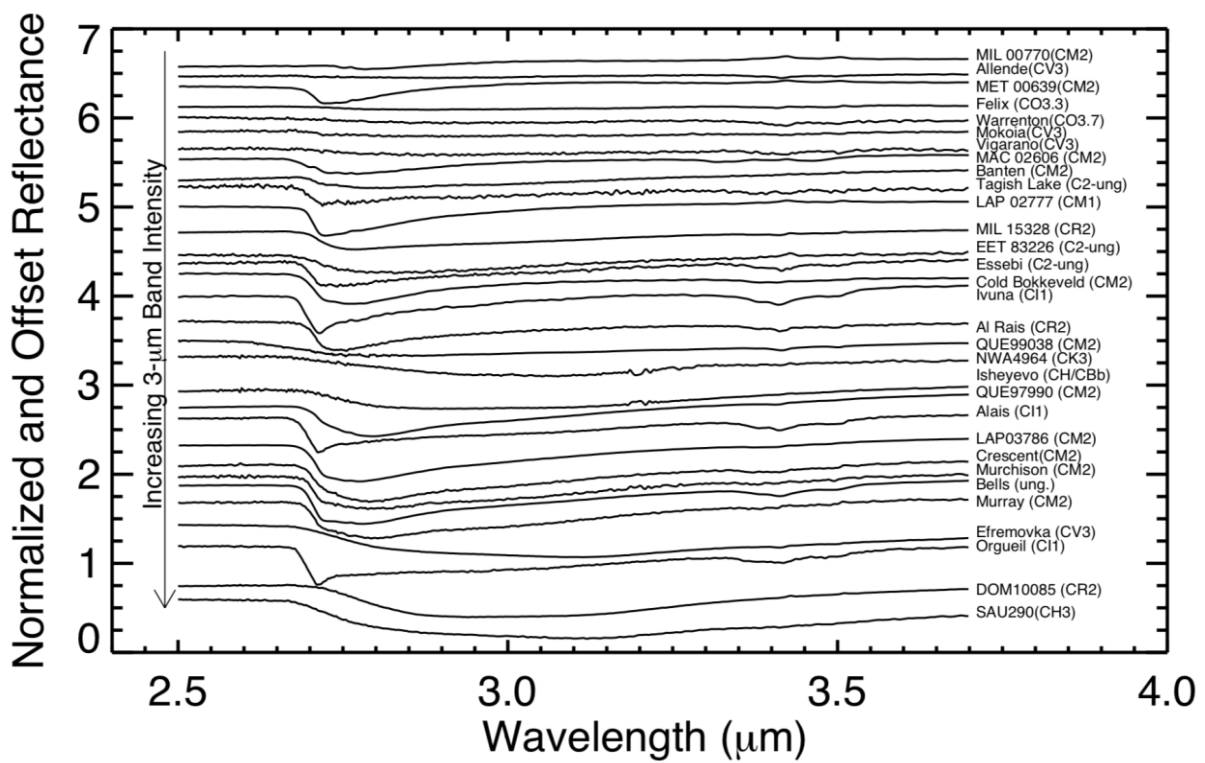

**Figure 5.** IR reflectance spectra of all studied carbonaceous chondrites covering all carbonaceous chondrite types available in terrestrial meteorite collections, including CI, CM, CR, CV, CO, CH, CB, CK, and C2 ungrouped chondrites. This figure also includes spectra of the 10 CM and CI chondrites studied in Takir et al. (2013). Spectra were measured under asteroid-like (vacuum and thermally-desiccated) conditions. The 3-μm band center increases

with increasing alteration. Band shape and center changes with degree of aqueous alterations.

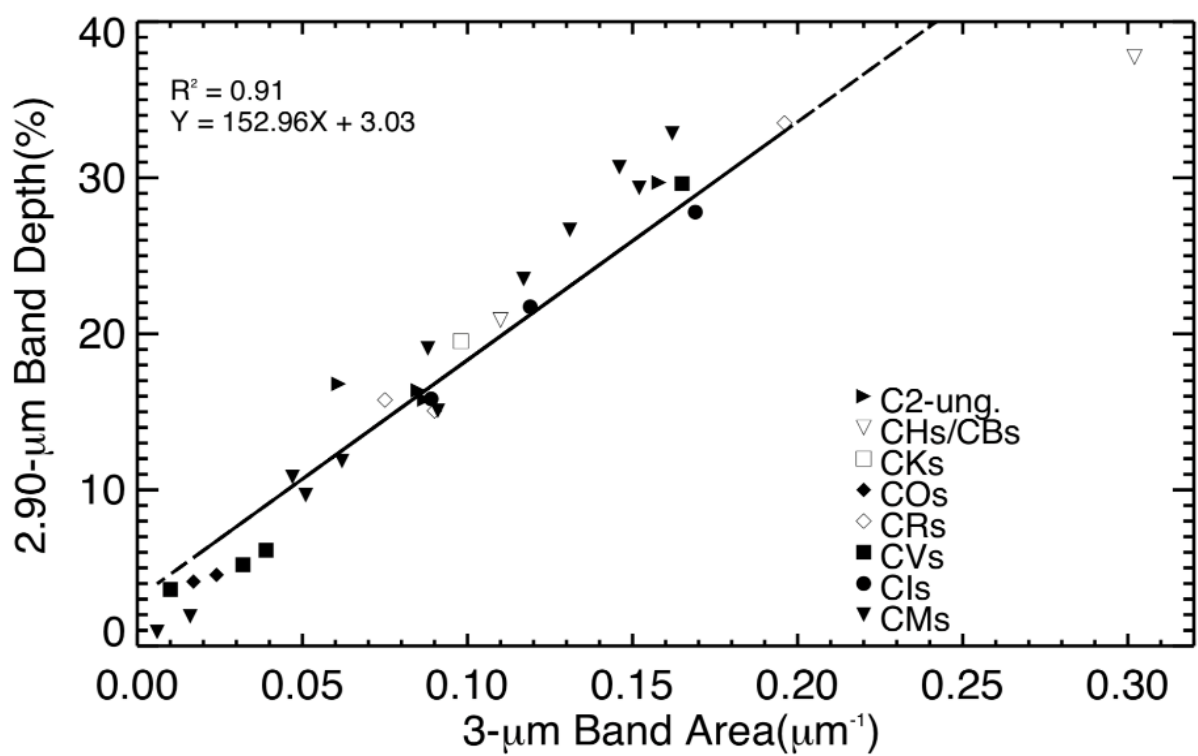

**Figure 6.** We found a good correlation between the 3-μm band area (BA) and 2.90-μm band depth ($BD_{2.90}$). In this plot, we combined data from this study and the study of Takir et al. (2013).

*CI carbonaceous chondrites*

        The 3-μm spectra of CI chondrites are distinct from all chondrite types' spectra with a deep and wide band in ambient conditions. This band became shallower, and the narrow shorter wavelength band (centered ~2.71 μm) is emphasized after driving off adsorbed water from the sample. We find the significant change between ambient and asteroid-like spectra in CI chondrites (due to the substantial adsorbed water) to be more likely associated with the presence of FeO as oxyhydroxides and complex clay minerals in these chondrites. Berlanga et al. (2016) noted a similar trend for CI chondrites after performing $CO_2$ gas adsorption experiments on various carbonaceous chondrite types. Additionally, CI chondrites spectrally show high abundances of organics, higher than any other carbonaceous chondrite type in this study. There is a possible association between the mineralogy of CI chondrites (oxyhydroxides- and clays-rich) and the high abundances of organics, which became more defined and pronounced in asteroid-like conditions (Figure 3).These results are supported by the findings of Pearson et al. (2002), who reported that organic material is

strongly associated with the phyllosilicate-rich matrix in carbonaceous chondrites.

The asteroid-like spectrum of Alais exhibits a very narrow 3-μm-band centered at ~2.71 μm, consistent with Ivuna (CI1) spectra measured by Takir et al. 2013. These authors found that Ivuna is highly altered and its spectra are characterized by 3-μm band centers at shorter wavelengths, consistent with the presence of expandable clays (e.g., saponite) and Mg-serpentine (e.g., lizardite, chrysotile). Spectra of Alais show that it underwent less aqueous alteration than Ivuna because Alais has a shallower 3-μm band. Endress and Bischoff (1996) also concluded that Alais experienced less aqueous alteration than Ivuna because of the lower abundance of carbonates in Alais.

Orgueil exhibits a similar 3-μm-band, centered ~2.71 μm, to the one found in Alais and Ivuna. However, the 3-μm-band in Orgueil is deeper, suggesting it experienced more aqueous alteration than Alais and Ivuna. This result is an agreement with Brearley and Martin (1992), who reported that Orgueil underwent more aqueous alteration than Alais and Ivuna on the basis of the depletion of Ca in their matrices. According to these authors, Ca is progressively leached from the matrix material to produce carbonate veins as a function of increased aqueous alteration. Endress and Bishoff (1996) also reported higher abundance of magnetite in Orgueil relative to Alais, providing more evidence that Orgueil experience more aqueous alteration.

*2.1. CM carbonaceous chondrites*

We found spectra of CM chondrites to have shallower 3-μm bands relative to CI chondrites, suggesting they experienced less aqueous alteration, which is in agreement with Takir et al. (2013). The authors found that CM chondrites are characterized by 3-μm band centers at longer wavelengths relative to Ivuna (CI1), consistent with the presence of Mg-serpentine (e.g., antigorite) and Fe-serpentine (cronstedtite), a serpentine for which $Fe^{3+}$ substitutes some Si in the tetrahedral site.

The 3-μm spectra of Murchison were found to be consistent with the CM chondrites studied in Takir et al. (2013). Murchison has a 3-μm band centered at longer wavelength (~2.80 μm), more likely due to the contribution of cronstedtite. Bunch and Cheng (1980) and Bland et al. (2014) also detected cronstedtite in Murchison using electron microprobe analysis and XRD analysis, respectively. Similarly, to Murchison, spectra of Crescent, Banten, and Murray exhibit a 3-μm band that is centered ~2.80 μm. Burns and Fisher (1991) found that cronstdetite (with $Fe^{2+}$ in the octahedral site and $Fe^{3+}$ in the octahedral and tetrahedral sites) proportions in Murray is lower than Murchison. Banten has a shallower 3-μm band, suggesting it experienced less aqueous alteration relative to Murchison, Crescent, and Murray. Using PSD-XRD analyses (the ratio of total phyllosilicate to total anhydrous Fe, Mg silicate), Howard et al. (2011) found that Murray experienced higher degree of aqueous alteration than Murchison. This is in agreement with the 3-μm-band area and depth calculated for Murray (BA=0.162 μm$^{-1}$, D$_{2.90}$= 30.60 %) and Murchison (BA=0.152 μm$^{-1}$, D$_{2.90}$ = 29.35%).

*2.2. CR carbonaceous chondrites*

Like CI and CM chondrites, CR chondrites are characterized by the presence of phyllosilicates. However, the abundances of phyllosilicates in most CR chondrites are lower relative to CI and CM chondrites (Van Schmus and Wood 1967, McSween 1977a, Weisberg et al. 1993). We measured 3-μm spectra of three CR chondrites: Al Rais, MIL 15328, and DOM 10085. The measured spectra of CR chondrites are consistent with a moderate degree of aqueous alteration in CR chondrites, which is thought to be moderate relative to CI and CM chondrites.

*2.3. CV carbonaceous chondrites*

We measured 3-μm spectra of 4 CV carbonaceous chondrites: two oxidized CVs, Allende and Mokoia, and two reduced CVs, Vigarano and Efremova. CV chondrites are thought to be affected by metamorphism (McSween 1977a, McSween 1979), providing important information about heating events of primitive materials in early Solar System. Using Mossbauer spectrometry of Allende and Vigarano, Bland et al (2002) concluded that oxidized CVs can be distinguished from reduced CVs by the higher volume of magnetite. Howard et al. (2015), however, found that the reduced vs. oxidized classification in CV chondrites may not be reflected in magnetite abundances. Using position sensitive detector X-ray diffraction (PSD-XRD) analyses, these authors found that metal abundances are greater in reduced CVs in which sulfides contains less Ni and can be used as a more effective discriminator between the two CV subgroups.

Except for Efremovka, we found all CV chondrites to have a 3-μm band that is very shallow (2.90-μm band depth of 4-6%), in agreement with Buseck and Hua (1993), who reported that phyllosilicate proportions in CV chondrites are very lower relative to the CI, CM, and CR chondrites. Efremovka, a reduced CV, is unique in that it has a broader 3-μm band (centered ~ 3.11 μm) and a deeper 2.90-μm band (~29%). This broad 3-μm band could be attributed to aqueous alteration and/or the presence of metal.

*2.4. CO Carbonaceous Chondrites*

We measured 3-μm spectra of two CO chondrites: Felix and Warrenton.

Like three of most CV chondrites in this study, Felix and Warrenton have very shallow and broad 3-μm band (4-5 %), centered ~2.90 μm, suggesting they experienced minor aqueous alteration. McSween (1977b) found that Warrenton is more metamorphosed than Felix. However, we found Felix's 3-μm band to be slightly shallower (BA = 0.017 μm$^{-1}$, D$_{2.90}$= 4.11 μm) compared with Warrenton's 3-μm band (BA=0.024 μm$^{-1}$, D$_{2.90}$= 4.56 μm).

*2.5. CK Carbonaceous Chondrites*

NWA 4964 was the only CK chondrite in this study (and one of the rarest carbonaceous chondrites in general) and is the only meteorite that shows significant absorptions at approximately 1 μm and 2 μm, which are due to Fe$^{2+}$ in the M1 and M2 crystallographic sites in olivine and pyroxene (Burns 1993). This indicates that NWA has a high fraction of anhydrous silicates. However, the high abundance of anhydrous silicates is not consistent with the 2.90-μm band depth of 19.53% that suggests NWA 4964 experienced a high degree of aqueous alteration and includes high phyllosilicates abundance. This discrepancy is possibly due to the moderate weathering grade of this dessert meteorite; the large amount of phyllosilicates and deep absorption band may be terrestrial alteration. This meteorite has a 3-μm spectrum similar to Efremovka, the only CV in this study that exhibits a broad feature centered ~3.06 μm with a 2.90-μm band depth of ~20%.

*2.6. CH/CBb Carbonaceous Chondrites*

Two 3-μm spectra of CH and CB chondrites (SAU 290 and Isheyevo) were measured in this study. SAU 290 was classified as a CH3 chondrite, which is characterized by high FeNi metal abundance and the presence of heavily hydrated lithic clasts (Krot et al. 2003). Isheyevo, on the other hand, includes several lithologies ranging from metal-poor to metal-rich. It was classified as a CH/CBb chondrite because of its metal-poor lithology that is texturally and mineralogically consistent with CH chondrites and its metal-rich lithology that resembles CBb's lithology (Ivanova et al. 2008). The 3-μm band in SAU 290 is unique in that it exhibits the deepest (D$_{2.90}$ = 38%) and broadest 3-μm band (centered ~3.11 μm) among all chondrites in this study. This is possibly because of the presence of heavily hydrated clasts and high abundance of FeNi metal in this chondrite. 3-μm spectra of Isheyevo also show a wide 3-μm band (centered ~2.90 μm), possibly due to FeNi metal, but shallower 3-μm band depth (D$_{2.90}$ = 21%) relative to SAU 290.

*2.7. C2-ung carbonaceous chondrites*

It is possible that some ungrouped chondrites in terrestrial meteorite collection belong to the eight established previously discussed in this paper. However, they have not been thoroughly analyzed to be given a definitive classification (Cloutis et 2012f). Here we measured 3-μm spectra of three ungrouped chondrites: EET 83226, Essebi, and Tagish Lake.

Essebi exhibits a 3-μm band that is very similar to Murchison and Murray, two CM chondrites. Howard et al. (2011) found that the C2-ung Essebi has lower degree of aqueous alteration than Murchison and Murray, using PSD-XRD analyses. This is in agreement with the 3-μm band parameters calculated for these chondrites. The 3-μm band in EET 83226 is broader, with a band center of 2.83 μm, and shallower relative to the 3-μm band in Essebi. This suggests that EET 83226 includes more Fe-serpentine and experienced less aqueous alteration. Tagish Lake was found to have four distinct lithologies: carbonate-poor, carbonate-rich, inclusion-poor magnetite- and sulfide-rich, and carbonate-rich, siderite-dominated (Zolensky et al., 2002, Izawa et al. 2010). In asteroid-like conditions, the 3-μm spectrum of Tagish Lake is characterized by a sharp band centered at ~2.72 μm, making it more consistent with CI chondrites. No strong carbonate $CO_3$ absorptions were found at the 3.4-3.5 μm and 3.8-4.0 μm regions in the studied Tagish Lake sample. Like CI chondrites, Tagish lake shows significant change between ambient and asteroid-like spectra (due to the substantial adsorbed water), more likely associated with the presence of FeO as oxyhydroxides and complex clay minerals.

**Summary and Conclusions**

We measured 3-μm reflectance spectra under asteroid-like conditions (vacuum and thermally-desiccated) of 21 chondrites covering all carbonaceous chondrite types available in terrestrial meteorite collections. With these high-vacuum laboratory spectroscopic measurements, we were able to simulate the asteroid environments and minimize the amount of adsorbed terrestrial water on samples, which affects spectral interpretations. Our study revealed that the 3-μm band in the spectra of the desiccated meteorites is diverse for the 21 studied carbonaceous chondrites, suggesting their parent bodies experienced distinct secondary process environments (aqueous alteration, metamorphism). We found an agreement between the 3-μm spectral characteristics of carbonaceous chondrites and carbonaceous chondrite classifications. These results are important for direct comparisons with and appropriate interpretation of reflectance spectra from ground-based telescopic and spacecraft observations.

CI chondrites have a very narrow 3-μm-band centered at ~2.71 μm, consistent with Mg-serpentine and clay minerals. CI chondrites were found to be affected by substantial amount of adsorbed water from laboratory atmospheric contamination that desorbed under vacuum and heating to ~ 375K for several hours. This adsorptive behavior is likely due to the high abundances of oxyhydroxides and complex clay minerals in these chondrites. In addition, we attribute the high abundances of organics revealed by 3-μm spectra of CI chondrites (higher than all other chondrite types) to the mineralogy of CI chondrites (oxyhydroxides- and clay-rich). CM chondrites have a less intense 3-μm band than CI chondrites, suggesting they experienced less aqueous alteration. The 3-μm band of CM chondrites is centered ~2.80 μm, more likely due to the presence of Fe-serpentine (cronstedtite). The measured 3-μm spectra of CR chondrites revealed that the degree of aqueous alteration in these chondrites is moderate (moderate phyllosilicate abundances) relative to CI and CM chondrites. CV chondrites, except for Efremovka, have a very shallow 3-μm band, consistent with their lower phyllosilicate proportions and their experienced metamorphism environments. Like CV chondrites, CO chondrites have a very shallow 3-μm band that suggest they experienced minor aqueous alteration. The 3-μm band in CH/CBb is deep and broad centered ~ 3.11 μm, possibly due the high abundance of FeNi metal and presence of heavily hydrated clasts in these chondrites. The 3-μm spectra of Essebi (C2-ung) and EET 83226 are more consistent with CM chondrites' spectra. 3-μm spectra of Tagish lake (C2-ung), on the other hand, are consistent with CI chondrites. These spectra show substantial change between ambient and asteroid-like conditions due to the significant amount of adsorbed water. This spectral change in Tagish Lake is more likely associated with the presence of oxyhydroxides and clays.

**Acknowledgements**

This work has been supported by NASA Hayabusa2 Participating Scientist grant NNX17AL02G (PI: DT). Travel support for YN to JHU APL's LabSPEC was provided by Japan Society for the Promotion of Science Core-to-Core Program "International Network of Planetary Sciences". We thank Devin Schrader for preparing and providing the following samples: MIL 15328, DOM 10085, and EET 83226. We thank Mike Zolensky for preparing and providing the following samples: NWA 4964, SAU 290, and Tagish Lake. The authors would also thank and appreciate Neyda Abreu and Kieren Howard for the carbonaceous chondrite discussion. Other meteorite samples have been provided by the American Museum of Natural History & University of Winnipeg's Hyperspectral Optical Sensing for Extraterrestrial Reconnaissance Laboratory. US Antarctic meteorite samples are recovered by the Antarctic Search for Meteorites (ANSMET) program, which has been funded by NSF and NASA, and characterized and curated by the Department of Mineral Sciences of the Smithsonian Institution and Astromaterials Curation Office at NASA Johnson Space Center.